\documentclass[5p]{elsarticle}

\usepackage{amsmath,amssymb,amsthm}
\usepackage{enumitem}
\usepackage{url}
\usepackage{booktabs}
\usepackage{xspace}
\usepackage{setspace}

\newcommand{\real}{\ensuremath{\mathbb{R}}}

\newcommand{\smat}[1]{\ensuremath{\left[\begin{smallmatrix}#1\end{smallmatrix}\right]}}
\newcommand{\bmat}[1]{\ensuremath{\begin{bmatrix}#1\end{bmatrix}}}

\newcommand{\tu}[1]{\textup{#1}}
\newcommand{\sa}[1]{\mathsf{#1}}

\DeclareMathOperator{\ve}{vec}

\DeclareMathSymbol{\cdoT}{\mathord}{symbols}{"01}
\DeclareMathOperator{\vol}{vol}

\newtheorem{lemma}{Lemma}

\newtheorem{proposition}{Proposition}

\newtheorem{example}{Example}

\allowdisplaybreaks
\graphicspath{{./figures/}}

\begin{document}

\title{Trade-offs in learning controllers from noisy data\tnoteref{t1}}
\tnotetext[t1]{This research is partially supported by a Marie Sk\l{}odowska-Curie COFUND grant, no.~754315 and by NWO, project no. 15472.}
\tnotetext[]{Corresponding author: Andrea Bisoffi.}
\author[1]{Andrea Bisoffi, Claudio De Persis%
}
\ead{\{a.bisoffi,c.de.persis\}@rug.nl}
\author[2]{Pietro Tesi}
\ead{pietro.tesi@unifi.it}

\address[1]{ENTEG and the J.C. Willems Center for Systems and Control, University of Groningen, 9747 AG Groningen, 
The Netherlands}
\address[2]{DINFO, University of Florence, 50139 Florence, Italy}

\makeatletter
\def\ps@pprintTitle{%
  \let\@oddhead\@empty
  \let\@evenhead\@empty
  \let\@oddfoot\@empty
  \let\@evenfoot\@oddfoot
}
\makeatother

\begin{abstract}
In data-driven control, a central question is how to handle noisy data. 
In this work, we consider the problem of designing a stabilizing controller for an unknown linear system using only a finite set of noisy data collected from the system. 
For this problem, many recent works have considered a disturbance model based on energy-type bounds. 
Here, we consider an alternative more natural model where the disturbance obeys instantaneous bounds.
In this case, the existing approaches, which would convert instantaneous bounds into energy-type bounds, can be overly conservative. 
In contrast, without any conversion step, simple arguments based on the S-procedure lead to a very effective controller design through a convex program. 
Specifically, the feasible set of the latter design problem is always larger, and the set of system matrices consistent with data is always smaller and decreases significantly with the number of data points.
These findings and some computational aspects are examined in a number of numerical examples.
\end{abstract}

\maketitle

\section{Introduction}
\label{sec:intro}

\subsection*{Problem formulation and related work}

In the last few years, there has been a renewed growing interest towards data-driven control. 
Data-driven control provides an alternative tool for control design whenever modeling based on first principles is difficult or impossible \cite{Sznaier_review,recht_annual} and it has been employed for robust and optimal control \cite{Baggio2019,depersis2020tac}, predictive control \cite{Coulson2018,Allibhoy2021}, and control of nonlinear \cite{Tabuada2020} and time-varying systems \cite{Nortmann2020}.

In this paper, we start from a basic control problem, that is, designing a controller by a finite set of data with $T$ samples collected from a system. 
Specifically, for state $x \in \real^n$, input $u \in \real^m$ and disturbance $d \in \real^n$, we apply an \textit{input sequence} $\{u(0),\dots,u(T-1)\}$ to a linear time-invariant discrete-time system and measure the \textit{state sequence} $\{x(0),\dots,x(T)\}$ generated as response as
\begin{equation}
\label{sys}
x(i+1) = A_\star x(i)+ B_\star u(i) + d(i), \quad i = 0, \dots, T-1.
\end{equation}
The matrices $A_\star$ and $B_\star$ are unknown and we do not have access to disturbance $d$, so data are \emph{noisy}. 
The objective is to use the input and (noisy) state sequence to design a feedback control law $u=Kx$ to solve a stabilization problem, that is, ensuring that $A_\star+B_\star K$ has eigenvalues with  magnitude strictly less than 1 (Schur stability).

Albeit basic, this problem poses many challenges. 
In fact, since $d$ in \eqref{sys} affects the state sequence, it is impossible to uniquely reconstruct the system matrices $A_\star$ and $B_\star$ from the input and state
sequences; instead, we have a \emph{set} of possible matrices that could have generated the state sequence.
Accordingly, the problem corresponds to a \emph{robust} stabilization problem in the face of the uncertainty introduced by $d$, entailing a set of matrices to stabilize rather than a singleton.
In the literature, several works have considered problems of this type, in the very same setting as here \cite{depersis2020tac,
berberich2019robust,berberich2019combining,vanwaarde2020noisy}, 
or in other settings like linear-quadratic-regulator design 
\cite{Dean2018journ,Xue2020,dpt2021Aut}, robust set invariance \cite{bisoffi2020controller},
switched and nonlinear systems \cite{Dai2018cdc,dai2021semi,Guo2020,Bisoffi2020}.

In all aforementioned works, the starting point is to define a model for the disturbance. Typical choices on opposite sides of the spectrum are disturbances with known probability distributions or statistics (e.g., $d$ is assumed a white Gaussian noise
\cite{Dean2018journ,Xue2020,dpt2021Aut})
or the so-called \emph{unknown}\emph{-}\emph{but}\emph{-}\emph{bounded} models, as we consider 
in this paper. In {unknown-but-bounded} models, $d$
may be a deterministic function of time or a stochastic process and
the only characteristic assumed to be known is that $d$ is contained in a given region
\cite{depersis2020tac,vanwaarde2020noisy,berberich2019robust,berberich2019combining,
dpt2021Aut,bisoffi2020controller,Dai2018cdc,Guo2020,Bisoffi2020}. 
In this setting, a natural way to describe the uncertainty is to consider \emph{instantaneous} constraints for $d$ as
\begin{equation}
\mathcal{D}_\tu{i} :=  \{ d \in \real^n \colon | d |^2 \le \epsilon\} \label{Di set}
\end{equation}
where $\epsilon \geq 0$ represents our prior knowledge on $d$.
This model naturally arises in many practical cases (e.g., with process load disturbances), and it has a long history in set-constrained control and set-membership identification \cite{Glover1971,Bertsekas1971,Hjalmarssoncdc}. 
As an alternative to instantaneous constraints, one can consider 
\emph{energy} constraints on $d$ as
\footnote{Throughout the paper, $\prec$ ($\preceq$) and $\succ$ ($\succeq$) denote, respectively, negative (semi)definiteness and positive (semi)definiteness for matrices. $I$ is the identity matrix. $\top$ denotes transpose.}
\begin{equation}
\mathcal{D}_\tu{e} := \big\{ d^0, d^1, \dots, d^{T-1} \in \real^n 
\colon  \! \sum_{i=0}^{T-1} d^i (d^i)^\top \preceq \epsilon_\tu{e} I \big\} \label{De set}
\end{equation}
where $d^0, d^1, \dots, d^{T-1}$ correspond to any disturbance sequence and $\epsilon_\tu{e} \ge 0$ represents our prior knowledge on the disturbance sequence. 
This model (or similar ones involving quadratic bounds on the possible disturbance sequences) has been considered in many recent works on data-driven control \cite{depersis2020tac,vanwaarde2020noisy,
berberich2019robust,berberich2019combining}. 
The reason for considering $\mathcal{D}_\tu{e}$ is essentially technical. By using $\mathcal{D}_\tu{e}$, the uncertainty
set of system matrices consistent with data is given by a single quadratic matrix inequality and this enables deriving simple stability conditions \cite[Thm. 5]{depersis2020tac}, or even necessary and sufficient conditions under a mild regularity property on data as in \cite[Thm. 14]{vanwaarde2020noisy}. 
From a practical viewpoint, however, this disturbance model 
is artificial. In fact, $\mathcal{D}_\tu{e}$ is typically 
built starting from $\mathcal{D}_\tu{i}$, i.e., by setting $\epsilon_\tu{e}=\epsilon T$; see 
\cite[\S IV]{berberich2019robust},
\cite[Ex. 3]{vanwaarde2020noisy} and \cite[\S VI]{berberich2019combining}. 
As a consequence, the bound $\epsilon_\tu{e} = \epsilon T$ in~\eqref{De set} increases with $T$ and this may lead to a set of data-consistent system matrices that grows with $T$ (see Example~\ref{ex:noMonoEx})
so that using larger data sets may even be detrimental, which is undesirable and counterintuitive.
Among the aforementioned recent works, few considered instantaneous bounds. They were used in \cite{Dai2018cdc,dai2021semi} for data-driven control of switched and polynomial systems, where 
the problem was reduced to a nonconvex quadratic program and then relaxed to a polynomial optimization problem. Working directly with $\mathcal{D}_\tu{i}$ was also pursued in~\cite[\S V]{berberich2019combining} as a secondary development, by using 
\emph{sum-of-squares} relaxations. 

\subsection*{Paper contribution}

In this paper, we investigate pros and cons of using directly an instantaneous bound instead of translating it into an energy bound. 
We show that working with $\mathcal{D}_\tu{i}$ 
is indeed advantageous   
since \textit{(i)} the uncertainty set resulting from $\mathcal{D}_\tu{i}$ is never larger than the one resulting from $\mathcal{D}_\tu{e}$ with $\epsilon_\tu{e}=\epsilon T$; \textit{(ii)} there exist simple design methods that work directly with $\mathcal{D}_\tu{i}$ and always return a stabilizing solution whenever the methods based on $\mathcal{D}_\tu{e}$ do. 
We also show that the extra numerical cost due to working with $\mathcal{D}_\tu{i}$ is quite modest up to large sets of data.
Compared to \cite{berberich2019combining}, our method rests on simpler arguments and uses design tools purely based on linear matrix inequalities (instead of polynomial constraint qualification conditions that may be hard to assess). 
As an auxiliary contribution, we introduce a notion of size for \emph{matrix} ellipsoids to numerically measure the uncertainty induced by $d$. This measure confirms that the uncertainty resulting from $\mathcal{D}_\tu{i}$ is typically order of magnitudes smaller than that resulting from $\mathcal{D}_\tu{e}$.

\subsection*{Structure}

Section~\ref{sec:prel} obtains a size notion for matrix ellipsoids. In Section~\ref{sec:controller probs}, we present the controller design problems for the energy-bound and instantaneous-bound approach, and relate them in terms of feasibility. In Section~\ref{sec:energy vs ampl:theory}, we further compare the two approaches by examining the corresponding sets of dynamical matrices consistent with data. The numerical investigation in Section~\ref{sec:num example} completes the comparison, and leads to the conclusions of Section~\ref{sec:concl}.

\section{Preliminaries}
\label{sec:prel}

In this section, we set up the notation and introduce a notion of size for matrix ellipsoids, which will be used in the numerical simulations of Section~\ref{sec:num example}.

\subsection{Notation} 

For a matrix $A$, $|A|$ denotes its induced 2-norm.
Given two matrices $A$ and $B$, $A \otimes B$ denotes their Kronecker product.
For a matrix $A=\bmat{a_1 & \cdots & a_n}\in  \real^{m \times n}$ partitioned according to its columns, $\ve(A):=[a_1^\top\,\,\, \cdots \,\,\,a_n^\top]^\top$ denotes its vectorization. 
A property of vectorization is that, for matrices $A,X,B$ and $C$ of compatible dimensions, the matrix equation $A X B = C$ is equivalent to $(B^\top \otimes A) \ve(X) = \ve(C)$ \cite[Lemma~4.3.1]{horn1991topics}.
For fixed natural numbers $m$ and $n$, the inverse of the vectorization operator takes as input a vector $a=[a_1^\top\,\,\, \cdots \,\,\,a_n^\top]^\top \in \real^{mn}$ partitioned in components $a_1, \dots, a_n \in \real^m$ and returns
\begin{equation}
\ve^{-1}_{m,n}(a):=\bmat{a_1 & \dots & a_n} \in  \real^{m \times n}.
\end{equation}
For matrices $A$, $B$ and $C$ of compatible dimensions, we abbreviate $A B  C (AB)^\top$ to $A B \cdoT C[\star]^\top$, where the dot in the second expression clarifies unambiguously that $AB$ are the terms to be transposed.

\subsection{A size notion for matrix ellipsoids}
\label{sec:sizes of matrix ell}

For symmetric matrices $\sa{P} \in \real^{p \times p}$, $\sa{Q} \in \real^{q \times q}$, $\sa{A} \in \real^{p \times p}$,  $\sa{C} \in \real^{q \times q}$ and matrices $\sa{Z}_\tu{c} \in \real^{p \times q}$, $\sa{B} \in \real^{p \times q}$, we term \emph{matrix ellipsoid} a set in one of the next two forms:%
\begin{subequations}
\label{ell matr}
\begin{align}
\mathcal{E}_\textup{mat} & :=  \{ \sa{Z} \in \real^{p \times q}  
\colon   (\sa{Z}-\sa{Z}_\tu{c} )^\top \sa{P}^{-2} (\sa{Z}-\sa{Z}_\tu{c} ) 
\preceq \sa{Q}\}, \label{ell matr P Zc Q} \\
\mathcal{E}_\textup{mat}' & := \{ \sa{Z} \in \real^{p \times q}  \colon
 \sa{Z}^\top \sa{A} \sa{Z} + \sa{Z}^\top \sa{B} +  \sa{B}^\top \sa{Z} 
 + \sa{C} \preceq 0 \} \label{ell matr A B C}
\end{align}
\end{subequations}
where
\begin{equation}
\label{ell matr assump}
\sa{P} \succ 0, \sa{Q} \succ 0 \text{ ~and~ } \sa{A}\succ 0, 
\sa{B}^\top \sa{A}^{-1} \sa{B} -\sa{C}\succ 0.
\end{equation}
The constraints $\sa{Q} \succ 0$ and $\sa{B}^\top \sa{A}^{-1} \sa{B} -\sa{C}\succ 0$ ensure that $\mathcal{E}_\textup{mat}$ and $\mathcal{E}_\textup{mat}'$ are not empty or do not reduce to a singleton; the constraints $\sa{P} \succ 0$ and $\sa{A}\succ 0$ ensure that the matrix ellipsoid is a bounded set.
We stress that many sets considered in the sequel (e.g., $\mathcal{C}$ in Section~\ref{sec:setsC} and $\overline{\mathcal{I}}$ in Section~\ref{sec:overapprox overline I set}) have to be expressed in terms of these matrix ellipsoids, 
and that \eqref{ell matr P Zc Q} and \eqref{ell matr A B C} are natural 
extensions of the classical ellipsoids in the Euclidean space, 
cf.~\cite[Eqs.~(3.8)-(3.9)]{boyd1994linear}.

Standard computations reformulate $\mathcal{E}_\tu{mat}'$ as
\begin{align}
\mathcal{E}_\tu{mat}' & = \{ \sa{Z} \in \real^{p \times q} \colon 
(\sa{Z}+\sa{A}^{-1} \sa{B} )^\top \sa{A} (\sa{Z}+\sa{A}^{-1} \sa{B} ) \nonumber  \\
& \hspace*{35mm} - (\sa{B}^\top \sa{A}^{-1} \sa{B} - \sa{C}) \preceq 0 \}. \label{ell matr A B C alt}
\end{align}
Hence, $\mathcal{E}_\tu{mat}$ and $\mathcal{E}_\tu{mat}'$ 
are the \emph{same} set for
\begin{equation}
\label{ell matr equiv}
\sa{Z}_\tu{c} = - \sa{A}^{-1} \sa{B} ,\, \sa{P}^{-2} = \sa{A},\, \sa{Q}=\sa{B}^\top \sa{A}^{-1} \sa{B} -\sa{C}.
\end{equation}
Let us focus on $\mathcal{E}_\tu{mat}$ to define a size. Since $\sa{Q} \succ 0$,  $\mathcal{E}_\tu{mat}$ is equivalently written as
\begin{align*}
\mathcal{E}_\tu{mat} & \!=\! \{ \sa{Z} \! \in \real^{p \times q} 
\colon \sa{Q}^{-1/2} (\sa{Z}-\sa{Z}_\tu{c} )^\top \sa{P}^{-2} 
(\sa{Z}-\sa{Z}_\tu{c} ) \sa{Q}^{-1/2} \! \preceq I \}\\
& \!= \!\{ \sa{Z}_\tu{c} + \sa{P} \sa{Y} \sa{Q}^{1/2} \colon \sa{Y} 
\in \real^{p \times q}, \sa{Y}^\top \sa{Y} \preceq I \}
\end{align*}
where $\sa{Y}^\top \sa{Y} \preceq I$ is equivalent to $|\sa{Y}| \le 1$.
Then, we specify the measure space in question and the adopted measure. 
Given natural numbers $p$ and $q$, 
$\ve\colon \real^{p \times q} \to \real^{p q}$ is a bijection 
between $\real^{p \times q}$ and $\real^{p q}$ with 
inverse $\ve^{-1}_{p,q} \colon \real^{p q} \to \real^{p \times q}$. 
With $\mathcal{P}(\mathcal{T})$ denoting the power set of a set $\mathcal{T}$, 
we define the bijection $\mathcal{V} \colon \mathcal{P}(\real^{p \times q}) \to \mathcal{P}(\real^{pq})$ as 
\begin{equation*}
\mathcal{V}(\mathcal S) := \{ \ve(s) \colon s \in \mathcal S \}
\end{equation*}
for $\mathcal S \subseteq \real^{p \times q}$. Then, its inverse $\mathcal{V}^{-1}$ satisfies 
$\mathcal{V}^{-1}(\mathcal S_\tu{v}) = \{\ve^{-1}_{p,q}(s_\tu{v}) 
\colon s_\tu{v} \in \mathcal S_\tu{v}\}$ for $\mathcal S_\tu{v} \subseteq \real^{pq}$. 
By using standard notions in measure theory \cite[\S 1.2,\,\S 1.4]{tao2011introduction}, 
we consider in $\real^{pq}$ the standard collection 
$\mathcal{B}_\tu{v}$ of Lebesgue measurable sets as 
$\sigma$-algebra and for $\beta_\tu{v} \in \mathcal{B}_\tu{v}$, 
$m(\beta_\tu{v})$ is the Lebesgue measure of the set $\beta_\tu{v}$. 
With $\mathcal{B}_\tu{v}$ and $m$, we can define a 
$\sigma$-algebra $\mathcal{B}$ in $\real^{p \times q}$ as 
$\mathcal{B}:=\{ \mathcal{V}^{-1}(\beta_\tu{v}) \colon \beta_\tu{v} \in \mathcal{B}_\tu{v} \}
$ and then, for each $\beta \in \mathcal{B}$, a measure $\mu$ as $\mu(\beta):= m(\mathcal{V}(\beta))$. 
This makes $(\real^{p \times q},\mathcal{B},\mu)$ a measure 
space \cite[Def.~1.4.27]{tao2011introduction}. For $\sa{z}_\tu{c} := \ve \sa{Z}_\tu{c}$, let
\begin{equation*}
\begin{split}
\mathcal{E}_\textup{vec} & :=  \mathcal{V}(\mathcal{E}_\tu{mat})\\
 & = \{ \sa{z}_\tu{c} + (\sa{Q}^{1/2} \otimes \sa{P}) \sa{y} \colon \sa{y} \in \real^{p q}, |\ve^{-1}_{p,q}(\sa{y}) | \le 1\},
 \end{split}
\end{equation*}
which is closed, and thus Lebesgue measurable 
\cite[Lemma 1.2.13]{tao2011introduction}. 
Accordingly, $\mathcal{E}_\tu{mat}\!$ belongs to $\mathcal{B}$ and its measure is
\begin{equation}
\label{measure of E mat}
\mu(\mathcal{E}_\tu{mat}) = m(\mathcal{E}_\tu{vec})=:\vol(\mathcal{E}_\tu{vec})
\end{equation}
where it is standard to identify the volume of $\mathcal{E}_\tu{vec}$ 
with its Lebesgue measure (see, e.g., \cite[p.~105]{boyd2004convex}). 
The next lemma determines $\vol(\mathcal{E}_\tu{vec})$.
\begin{lemma}
\label{lemma:size vectorized}
Let $p$ and $q$ be given natural numbers, 
and let $\sa{Q}\in \real^{q \times q}$ and $\sa{P} \in \real^{p \times p}$ be symmetric positive definite matrices. Then, 
\begin{equation*}
\vol(\mathcal{E}_\textup{vec}) = \beta (\det \sa{Q})^\frac{p}{2} (\det \sa{P})^{q}
\end{equation*}
where $\beta$ is a constant that depends only on $p$ and $q$.
\end{lemma}
\begin{proof}
Let 
\begin{equation*}
\overline{\mathcal{E}}_\textup{vec} :=  
\{ (\sa{Q}^{1/2}  \otimes \sa{P}) \sa{y} \colon 
\sa{y} \in \real^{p q}, |\ve^{-1}_{p,q}(\sa{y}) | \le 1\}.
\end{equation*} 
By translation invariance \cite[Exercise 1.2.20]{tao2011introduction}, $m(\overline{\mathcal{E}}_\tu{vec}) = m(\mathcal{E}_\tu{vec})$. 
The set $\{ \sa{y} \in \real^{p q} \colon |\ve^{-1}_{p,q}(\sa{y}) | \le 1\}$
is closed, thus Lebesgue measurable. By \cite[Exercise 1.2.21]{tao2011introduction},
\begin{equation*}
m( \overline{\mathcal{E}}_\textup{vec})  = \det ( \sa{Q}^{1/2}  \otimes \sa{P}) m\big(   
\{ \sa{y} \in \real^{p q} \colon |\ve^{-1}_{p,q}(\sa{y}) | \le 1\}
\big). 
\end{equation*}
The last term in this expression depends only on $p$ and $q$ and  
$\det ( \sa{Q}^{1/2}  \otimes \sa{P}) = (\det \sa{Q})^\frac{p}{2} (\det \sa{P})^q$ by Kronecker-product properties \cite[\S 4.2, Problem 1]{horn1991topics}.
\end{proof}

In view of \eqref{measure of E mat} and 
Lemma \ref{lemma:size vectorized}, the set $\mathcal{E}_\tu{mat}$ has measure 
$\mu(\mathcal{E}_\tu{mat})= \beta (\det \sa{Q})^\frac{p}{2} (\det \sa{P})^{q}$. 
The constant $\beta$ is analogous 
to the proportionality constant in \cite[p. 42]{boyd1994linear}, 
which is the volume of the $2$-norm unit ball. 
The precise expression of $\beta$ is irrelevant for our developments because we will compare only matrix ellipsoids with  same $p$ and $q$. Hence, we disregard $\beta$ and define the \emph{size} of $\mathcal{E}_\textup{mat}$ simply as
\begin{equation*}
(\det \sa{Q})^{\frac{p}{2}}(\det \sa{P})^{q}.
\end{equation*}
Since the sets $\mathcal{E}_\textup{mat}$ and $\mathcal{E}_\textup{mat}'$ are the same when the correspondences in~\eqref{ell matr equiv} hold, their size is in that case
\begin{equation}
\label{size n>=1}
\hspace*{-1mm}(\det \sa{Q})^{\frac{p}{2}}(\det \sa{P})^{q} \!= \! (\det (\sa{B}^\top \sa{A}^{-1} \sa{B} -\sa{C}))^{\frac{p}{2}}(\det (\sa{A}^{-1}))^{\frac{q}{2}}.
\end{equation}
Both terms in this expression reduce to classical volume formulae for $\sa{Z} \in \real^{p \times 1}$ \cite[p.~42]{boyd1994linear}.

\section{Data-consistent dynamics and controller design problems}
\label{sec:controller probs}

In this section we return to our original problem of designing a stabilizing controller for \eqref{sys} from a finite data set.
We first determine the set of system matrices consistent 
with the data points when the disturbance model is given by $\mathcal{D}_\textup{i}$ or $\mathcal{D}_\textup{e}$, and then formulate the corresponding control design problems. 
From~\eqref{Di set}, \eqref{De set} and $\epsilon_\tu{e}=\epsilon T$ as discussed in Section~\ref{sec:intro}, we have
\begin{subequations}
\begin{align}
\hspace*{-1mm}\mathcal{D}_\textup{i} = & 
\{ d \in \real^n \colon |d|^2 \le \epsilon \} =
\{ d \in \real^n \colon d d^\top \preceq \epsilon I \} \label{Di set eps}\\
\hspace*{-1mm}\mathcal{D}_\textup{e} = & \Big\{ d^0, \dots, d^{T-1}\in \real^n \colon T \epsilon I - 
\sum_{i=0}^{T-1} d^i (d^i)^\top   \nonumber \\
& \hspace*{0.8mm}
=\left[
\begin{array}{c|ccc}
\!\! I \! & d^0 \!\!\! & \dots \!\!\! & d^{T-1} \!\!
\end{array}
\right] \! \cdoT \!
\bmat{T\epsilon I & 0\\ 0 & -I}
[\star]^\top
\! \succeq 0 \Big\}.
\label{De set eps}
\end{align}
\end{subequations}
For notational convenience, all data points are grouped as
\begin{align*}
X_1 & :=\bmat{x(1) & x(2) & \cdots & x(T)}\\
X_0 & :=\bmat{x(0) & x(1) & \cdots & x(T-1)}\\
U_0 & :=\bmat{u(0) & u(1) & \cdots & u(T-1)}
\end{align*}
and the set of their relevant indices $i$ is $\mathbb{I}:=\{0,\dots, T-1\}$.

\subsection{Dynamics consistent with the data} 
\label{sec:setsC}

We call \emph{consistent with data} all the matrices $(A,B)$ that, for the selected input sequence, could have generated the measured state sequence while keeping $d$ in the bound $\mathcal{D}_\textup{e}$ in~\eqref{De set eps} or $\mathcal{D}_\textup{i}$ in~\eqref{Di set eps}, and we characterize the corresponding two sets in this section. 

Based on the bound in~\eqref{De set eps}, 
the matrices $(A,B)$ consistent with the data points are in
\begin{equation}
\begin{split}
& \mathcal{C} := \Big\{ (A, B) \colon D \in \real^{n \times T},
X_1 = A X_0 + B U_0 + D, \\
& \hspace*{25mm}\bmat{I &  D}
\bmat{T\epsilon I & 0\\ 0 & -I}
\bmat{I \\ D^\top}
\succeq 0
\Big\},
\end{split}
\label{C set}
\end{equation}
i.e., all those matrices for which some disturbance sequence satisfying the bound in~\eqref{De set eps} could have generated the measured data. 
By eliminating $D$ in \eqref{C set}, $\mathcal{C}$ rewrites as
\begingroup
\setlength\arraycolsep{2.5pt}
\thinmuskip=0mu plus 1mu
\medmuskip=0mu plus 2mu
\thickmuskip=1mu plus 3mu
\begin{equation}
\hspace*{-1.mm}\mathcal{C} = \Big\{ (A, B) \colon 
\bmat{I & A & B}
\bmat{I & X_1 \\ 0 & -X_0 \\ 0 & -U_0} \cdoT
\bmat{T\epsilon I & 0\\ 0 & -I}
[\star]^\top \succeq 0
\Big\}.\hspace*{-1mm}
\label{C set alt}
\end{equation}
\endgroup
Analogously, based on the bound in~\eqref{Di set eps}, the matrices $(A,B)$ consistent with a data point $i \in \mathbb{I}$ are in
\begin{equation}
\label{Ci set}
\begin{split}
& \mathcal{C}_i := \Big\{
(A,B) \colon  d \in \real^n, \\ 
& \hspace*{12mm} x(i+1) = A x(i) + B u(i) + d,\,
 d d^\top \preceq \epsilon I
\Big\},
\end{split}
\end{equation}
i.e., all those matrices for which some disturbance 
$d$ satisfying the bound in~\eqref{Di set eps} could have generated the measured data point $i$. 
By eliminating $d$ in \eqref{Ci set}, $\mathcal{C}_i$ rewrites as
\begingroup
\setlength\arraycolsep{2.5pt}
\thinmuskip=0mu plus 1mu
\medmuskip=0mu plus 2mu
\thickmuskip=1mu plus 3mu
\begin{equation}
\hspace*{-2mm}
\mathcal{C}_i  =  \Big\{ (A, B) \colon 
\bmat{I & A & B}
\bmat{I & x(i+1)\\ 0 & -x(i) \\ 0 & -u(i)}  \cdoT 
\bmat{\epsilon I & 0\\ 0 & -I}
[\star]^\top \succeq 0
\Big\}.\hspace*{-1mm} \label{Ci set alt}
\end{equation}
\endgroup
The expression of $\mathcal{C}_i$ in~\eqref{Ci set alt} 
mirrors that of $\mathcal{C}$ in~\eqref{C set alt}, 
which was the reason to write $d d^\top \preceq \epsilon I$ 
in~\eqref{Ci set} instead of the equivalent but more immediate $|d|^2 \le {\epsilon}$.
The set of matrices $(A,B)$ consistent with all data points is then
\begin{equation}
\label{inter Ci set}
\mathcal{I}:=\bigcap_{i=0}^{T-1} \mathcal{C}_i.
\end{equation}
When neglecting the constraint $\sa{A}\succ 0$ in~\eqref{ell matr assump}, both the sets $\mathcal{C}$ and $\mathcal{C}_i$ can be
expressed in the form \eqref{ell matr A B C} of matrix ellipsoids.
Indeed, the inequalities constituting the sets $\mathcal{C}$ in~\eqref{C set alt} and $\mathcal{C}_i$ in~\eqref{Ci set alt} can be equivalently written as
\begin{subequations}
\label{C Ci set ABC}
\begin{equation}
\bmat{A & B} \sa{A}_j \bmat{A^\top\\ B^\top} + \bmat{A & B} \sa{B}_j +
\sa{B}_j^\top \bmat{A^\top\\ B^\top} + \sa{C}_j \preceq 0
\end{equation}
where the matrices $\sa{A}_j, \sa{B}_j$, $\sa{C}_j$ with 
$j=\mathbb{I}$ for $\mathcal{C}$ and $j=i$ for $\mathcal{C}_i$
are defined in~\eqref{ABC_C_Ci}, displayed below over two columns. 
This corresponds to \eqref{ell matr A B C} with $\sa{Z}^\top=\bmat{A & B}$.
We note two facts. The set $\mathcal{C}$ is not necessarily bounded. It is bounded if $\sa{A}_\mathbb{I} \succ 0$ or, equivalently, the matrix $\smat{X_0 \\ U_0}$ has full row rank.
This condition expresses the property that data are sufficiently rich in content, and is related to the notion
of \emph{persistency of excitation} \cite{willems2005note}.
On the other hand, condition $\sa{A}_i \succ 0$ for $\mathcal{C}_i$ is never satisfied. However, this is irrelevant since the set of interest when dealing with instantaneous bounds is $\mathcal I$,  formed by the \emph{intersection} of all $\mathcal{C}_i$ as in \eqref{inter Ci set}. 
We will prove below in Proposition \ref{prop:set I included in set C} that $\mathcal I \subseteq \mathcal C$, which guarantees that $\mathcal I$ is bounded whenever $\mathcal C$ is.%
\begin{figure*}[h!]%
\begingroup%
\setlength\arraycolsep{2.5pt}%
\thinmuskip=0mu plus 1mu
\medmuskip=0mu plus 2mu
\thickmuskip=1mu plus 3mu
\begin{small}%
\begin{equation}%
\label{ABC_C_Ci}
\hspace*{-2mm}
\sa{C}_\mathbb{I} :=-T \epsilon I +X_1 X_1^\top\hspace*{-2pt},~
\sa{B}_\mathbb{I} := -\bmat{X_0 \\ U_0} X_1^\top\hspace*{-2pt},~
\sa{A}_\mathbb{I} := \bmat{X_0 \\ U_0}\bmat{X_0 \\ U_0}^\top\hspace*{-2pt},~
\sa{C}_i :=-\epsilon I +x(i+1)x(i+1)^\top\hspace*{-2pt},~
\sa{B}_i := -\bmat{x(i)\\u(i)}x(i+1)^\top\hspace*{-2pt},~
\sa{A}_i := \bmat{x(i)\\u(i)}\bmat{x(i)\\u(i)}^\top
\hspace*{-2mm}
\end{equation}%
\end{small}%
\endgroup
\hrulefill
\end{figure*}%
\end{subequations}%

\subsection{Controller design methods}

We now formulate the control design problems related to
the two sets $\mathcal{C}$ and $\mathcal{I}$, i.e., to 
the energy bound $\mathcal{D}_\textup{e}$ and to the instantaneous bound $\mathcal{D}_\textup{i}$.

The energy-bound approach solves:
\begin{align}
& \text{find} & & P\succ 0, K \nonumber \\
& \text{s.t.} & & (A + B K) P (A + B K)^\top -P \prec 0  \label{energy-based approach}\\
& & & \hspace*{7mm} \text{ for all } (A,B) \text{ such that } (A,B) \in \mathcal{C}
\nonumber
\end{align}
where the condition in the second line expresses discrete-time asymptotic stability.
The instantaneous-bound approach solves:
\begin{align}
& \text{find} & & P\succ 0, K \nonumber \\
& \text{s.t.} & & (A + B K) P (A + B K)^\top - P \prec 0   \label{ampl-based approach} \\
& & & \hspace{7mm}\text{ for all } (A,B) \text{ such that } (A,B) \in \mathcal{I}. \nonumber
\end{align}
Both the feasibility problems \eqref{energy-based approach} 
and \eqref{ampl-based approach} correspond to robust stabilization 
in the face of the uncertainty introduced by $d$, 
which induces a set of matrices to stabilize rather than a singleton. 
We stress that we consider here a \emph{stabilization} problem as a prototypical control problem, but our main conclusions would apply unchanged to other control problems 
like $\mathcal{L}_2$ gain minimization or $\mathcal{H}_\infty$ control.

By the lossless matrix S-procedure \cite[Thm. 12]{vanwaarde2020noisy} and Schur complement, the result \cite[Thm. 13]{vanwaarde2020noisy} shows that, under a mild regularity condition on data, feasibility of \eqref{energy-based approach} is equivalent to feasibility of

\begin{small}%
\vspace*{-12pt}
\begin{align}
& \text{find} & & P, Y, \beta, \alpha \nonumber \\
& \text{s.t.} & &
\left[\begin{array}{cccc}
P-\beta I & 0 & 0 & 0 \\
0 & -P & -Y^{\top} & 0 \\
0 & -Y & 0 & Y \\
0 & 0 & Y^{\top} & P
\end{array}\right] \label{energy-based approach <=>} \\
& & & \hspace*{1mm} -\alpha\left[\begin{array}{cc}
I & X_{1} \\
0 & -X_{0} \\
0 & -U_{0} \\
0 & 0
\end{array}\right]\left[\begin{array}{cc}
T \epsilon I & 0 \\
0 & -I
\end{array}\right]\left[\begin{array}{cc}
I & X_{1} \\
0 & -X_{0} \\
0 & -U_{0} \\
0 & 0
\end{array}\right]^{\top} \succeq 0 \nonumber\\
& & & \alpha \ge 0, \beta > 0, P \succ 0 \nonumber
\end{align}
\end{small}%
where now the decision variables appear linearly. 
If \eqref{energy-based approach <=>} has 
a solution, $K=Y P^{-1}$ is a stabilizing controller for~\eqref{sys}.

On the other hand, a \emph{tractable} equivalent reformulation 
of~\eqref{ampl-based approach} based on matrix ellipsoids cannot be obtained. 
Indeed, even in the simplest case where $n=m=1$, the set $\mathcal{I}$ 
is an intersection of ellipsoids and finding the ellipsoid with 
minimum volume containing $\mathcal{I}$ is NP-complete 
\cite[p.~44]{boyd1994linear}. Hence, finding such an 
ellipsoid and applying then the necessary and sufficient conditions in \cite[Thm. 12]{vanwaarde2020noisy} is impractical. To pursue an alternative approach, we need the next lemma, which is an immediate extension to matrices of the classic lossy S-procedure \cite[\S 2.6.3]{boyd1994linear}.
\begin{lemma}
\label{lemma:lossy S-proc}
Let $T_0$, \dots, $T_\ell \in \real^{(q+p) \times (q+p)}$ be symmetric matrices. If there exist \emph{nonnegative} scalars $\tau_1,\dots,\tau_\ell$ such that $T_0 - \sum_{i=1}^{\ell} \tau_i T_i \succeq 0$, then $\smat{I\\ Z}^\top T_0 \smat{I\\ Z} \succeq 0$ for all $Z\in \real^{p \times q}$ such that $\smat{I\\ Z}^\top T_i \smat{I\\ Z} \succeq 0$ for each $i = 1,\dots, \ell$.
\end{lemma}
\begin{proof}
For an arbitrary $Z$ satisfying $\smat{I\\ Z}^\top T_i \smat{I\\ Z} \succeq 0$ for each $i = 1,\dots, \ell$, use the condition $T_0 - \sum_{i=1}^{\ell} \tau_i T_i \succeq 0$ to obtain $\smat{I\\ Z}^\top T_0 \smat{I\\ Z} \succeq 0$.
\end{proof}

We use the lossy matrix 
S-procedure in Lemma~\ref{lemma:lossy S-proc} and obtain 
a sufficient condition to guarantee feasibility 
of~\eqref{ampl-based approach} in the next proposition.
\begin{proposition}
\label{prop:ampl-based approach <=}
Feasibility of
\begin{small}
\begin{align}
& \text{find} & & \hspace{-2mm} P, Y, \beta, \tau_0, \dots, \tau_{T-1} \nonumber \\
& \text{s.t.} & & \hspace{-2mm}
\left[\begin{array}{@{}cccc@{}}
P-\beta I & 0 & 0 & 0 \\
0 & -P & -Y^{\top} & 0 \\
0 & -Y & 0 & Y \\
0 & 0 & Y^{\top} & P
\end{array}\right] \label{ampl-based approach <=} \\
& & & \hspace*{1mm} 
\hspace{-2mm} - \sum_{i=0}^{T-1} \tau_i
\left[\begin{array}{@{}cc@{}}
I & x(i+1) \\
0 & -x(i) \\
0 & -u(i) \\
0 & 0
\end{array}\right]\left[\begin{array}{@{}cc@{}}
\epsilon I & 0 \\
0 & -I
\end{array}\right]\left[\begin{array}{@{}cc@{}}
I & x(i+1) \\
0 & -x(i) \\
0 & -u(i) \\
0 & 0
\end{array}\right]^{\top} \succeq 0 \nonumber\\
& & & \hspace{-2mm}\beta > 0,\tau_0\ge 0, \dots, \tau_{T-1}\ge 0, P \succ 0, \nonumber
\end{align}
\end{small}%
implies feasibility of \eqref{ampl-based approach}.
\end{proposition}
\begin{proof}
Condition $(A + B K) P (A + B K)^\top - P\prec 0$ 
in \eqref{ampl-based approach} is equivalent 
to the existence of a positive scalar $\beta$ such that 
$(A + B K) P (A + B K)^\top - P \preceq -\beta I$, i.e.,
\begin{small}
\begin{equation*}
\bmat{I & A & B} \cdoT
\bmat{
P-\beta I & 0 & 0 \\
0 & -P & -P K^{\top} \\
0 & -K P & - K P K^\top }[\star]^\top \succeq 0.
\end{equation*}
\end{small}%
By Schur complement and change of variable 
$Y= KP$, \eqref{ampl-based approach <=} is equivalent to
\begin{small}
\begin{align*}
& \text{find} & & \hspace{-2mm} P, K, \beta, \tau_0, \dots, \tau_{T-1} \nonumber \\
& \text{s.t.} & & \hspace{-2mm}
\left[\begin{array}{@{}cccc@{}}
P-\beta I & 0 & 0 \\
0 & -P & -P K^{\top} \\
0 & -K P & - K P K^\top \\
\end{array}\right] \label{ampl-based approach <=} \\
& & & \hspace*{1mm} 
\hspace{-2mm} - \sum_{i=0}^{T-1} \tau_i
\left[\begin{array}{@{}cc@{}}
I & x(i+1) \\
0 & -x(i) \\
0 & -u(i) \\
\end{array}\right]\left[\begin{array}{@{}cc@{}}
\epsilon I & 0 \\
0 & -I
\end{array}\right]\left[\begin{array}{@{}cc@{}}
I & x(i+1) \\
0 & -x(i) \\
0 & -u(i) \\
\end{array}\right]^{\top} \succeq 0 \nonumber\\
& & & \hspace{-2mm}\beta > 0,\tau_0\ge 0, \dots, \tau_{T-1}\ge 0, P \succ 0. \nonumber
\end{align*}
\end{small}%
By Lemma~\ref{lemma:lossy S-proc}, 
feasibility of this problem implies feasibility of~\eqref{ampl-based approach}, which proves the statement.
\end{proof}

As in~\eqref{energy-based approach <=>}, the decision variables 
appear linearly in~\eqref{ampl-based approach <=} and solving \eqref{ampl-based approach <=} yields the controller gain $K=Y P^{-1}$. With respect 
to~\eqref{energy-based approach <=>}, solving \eqref{ampl-based approach <=} 
involves $T$ scalar variables $\tau_i$ instead of a single one $\alpha$. 
Having as many decision variables as the number 
of data points may be computationally demanding with (very) large data sets. We will further elaborate on this point later 
in Section~\ref{sec:num example:unstable}. 

The feasibility properties of \eqref{energy-based approach <=>} 
and \eqref{ampl-based approach <=} are related. In particular, the
next proposition shows
that solving \eqref{ampl-based approach <=} is as easy or 
easier than solving \eqref{energy-based approach <=>}. 

\begin{proposition}
\label{prop:feas implication}
If \eqref{energy-based approach <=>} is feasible, \eqref{ampl-based approach <=} is feasible.
\end{proposition}

\begin{proof} Consider a feasible solution to 
\eqref{energy-based approach <=>} for some variables $P$, $Y$, 
$\beta$, $\alpha$, and note that
\begin{small}
\begingroup
\setlength\arraycolsep{2.pt}
\begin{equation*}
\begin{split}
\bmat{
I & X_{1} \\
0 & -X_{0} \\
0 & -U_{0} \\
0 & 0
} \cdoT
\bmat{
T \epsilon I & 0 \\
0 & -I
}
[\star]^{\top}
=
\sum_{i=0}^{T-1}
\bmat{
I & x(i+1) \\
0 & -x(i) \\
0 & -u(i) \\
0 & 0
}\cdoT
\bmat{
\epsilon I & 0 \\
0 & -I
}
[\star]^{\top}
\end{split}.
\end{equation*}
\endgroup
\end{small}%
Thus, the same variables $P$, $Y$, $\beta$ with 
$\tau_0=\dots=\tau_{T-1}=\alpha \ge 0$ provide a feasible solution to~\eqref{ampl-based approach <=}.
\end{proof}

In the next section, we aim at quantifying the gap in terms of feasibility. 
To this end, we shift the focus from the design problems 
\eqref{energy-based approach} and \eqref{ampl-based approach} to 
the corresponding sets $\mathcal{C}$ and $\mathcal{I}$ because 
\eqref{energy-based approach} and \eqref{ampl-based approach} 
impose the very same stability condition on $\mathcal{C}$ and $\mathcal{I}$, respectively. 
Hence, the smaller $\mathcal{C}$ or $\mathcal{I}$ is, the larger 
the \emph{feasible set} \cite[\S 4.1.1]{boyd2004convex} of the 
design problems \eqref{energy-based approach} 
or \eqref{ampl-based approach} is.
Next section will actually show that $\mathcal{I}\subseteq \mathcal{C}$.

\section{Comparison between energy and instantaneous bounds: Theoretical evidence}
\label{sec:energy vs ampl:theory}

In this section, we firstly show a monotonicity property of the set 
$\mathcal{I}$ with respect to the number $T$ of data points, 
and, secondly, a relation between the sets $\mathcal{C}$ and $\mathcal{I}$.

Firstly, to illustrate monotonicity properties, we would like to highlight the dependence of $\mathcal{C}$ 
and $\mathcal{I}$ on the number $T$ of data points. 
With some notation abuse, we then write equivalently 
$\mathcal{C}$ or $\mathcal{C}(T)$, and $\mathcal{I}$ or $\mathcal{I}(T)$.
It is immediate that%
\begin{equation}
\mathcal{I}(T+1) \subseteq \mathcal{I}(T)
\end{equation}
because $\mathcal{I}(T+1)$ contains precisely the same 
constraints as $\mathcal{I}(T)$ plus an additional one. 
This is a desirable property because we expect 
more data to reduce the uncertainty by providing additional 
information, or, in the worst case, to keep the uncertainty at the same level. 
On the other hand, we build the next example to show 
that $\mathcal{C}(T+1) \subseteq \mathcal{C}(T)$ does not hold in general.

\begin{figure}
\centerline{\includegraphics[scale=.85]{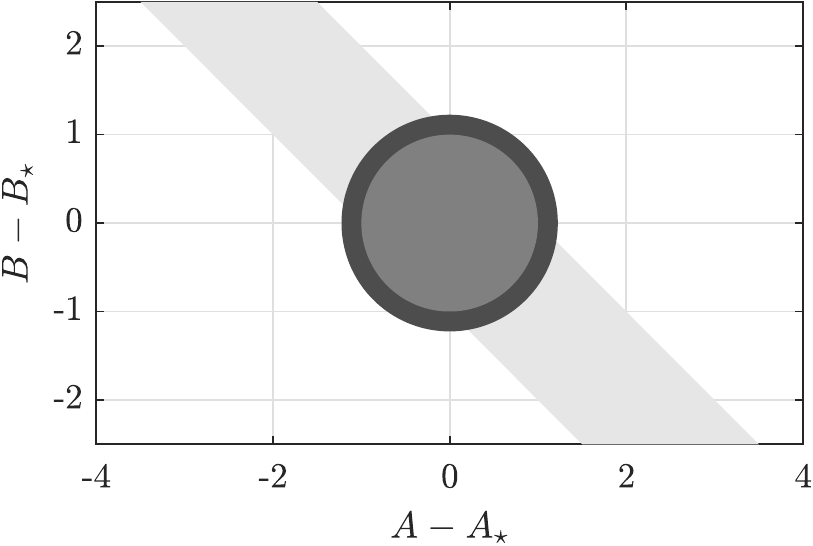}}
\caption{The sets $\mathcal{C}(1)$, $\mathcal{C}(2)$, $\mathcal{C}(3)$ defined in Example~\ref{ex:noMonoEx} correspond respectively to light, medium, dark gray.}
\label{fig:noMonoEx}
\end{figure}

\begin{example}
\label{ex:noMonoEx}
Consider in \eqref{sys} $A_\star=1/2$ and $B_\star=1/2$ as the unknown dynamical matrices.
Assume that $x(0)=1$, $u(0)=1$, $u(1)=-1$, $u(2)=0$, 
$d(0)=d(1)=d(2)=0$, and that the disturbance bound is $\epsilon=1$, 
These sequences give $x(1)=1$, $x(2)=x(3)=0$. 
Define for brevity $\tilde A := A - A_\star$ and 
$\tilde B := B - B_\star$. With some computations, \eqref{C set alt} 
yields for $T=1, 2, 3$
\begin{align*}
\mathcal{C}(1)   & = \{ (A, B) \colon 1 - \tilde A^2 - \tilde B^2 - 2\tilde A \tilde B \ge 0 \}\\
\mathcal{C}(2)   & = \{ (A, B) \colon 2 - 2\tilde B^2 - 2\tilde A^2 \ge 0\}\\
\mathcal{C}(3)   & = \{ (A, B) \colon 3 - 2\tilde B^2 - 2\tilde A^2 \ge 0 \},
\end{align*}
which are depicted in Fig.~\ref{fig:noMonoEx}. 
The ellipsoid corresponding to $\mathcal{C}(T)$ does not always 
shrink with larger $T$, but more data (such as when $T$ goes from $2$ to $3$) 
can actually induce larger bounds in an undesirable way. 
\end{example}

Secondly, the next proposition provides a relation between $\mathcal{C}$ and $\mathcal{I}$, which corroborates Proposition \ref{prop:feas implication}.
\begin{proposition}
\label{prop:set I included in set C}
The relation $\mathcal{I} \subseteq \mathcal{C}$ holds.
\end{proposition}
\begin{proof}We assume that $(A,B)$ belongs to 
$\mathcal{I}=\bigcap_{i=0}^{T-1} \mathcal{C}_i$, 
and show that, then, $(A,B)$ belongs to $\mathcal{C}$. 
Let us then manipulate the left-hand side of the inequality in \eqref{C set alt} as
\begin{small}
\begingroup
\setlength\arraycolsep{2.5pt}
\begin{align*}
& \bmat{I & A & B}
\bmat{I & X_1\\ 0 & -X_0 \\ 0 & -U_0} \cdoT
\bmat{T\epsilon I & 0\\ 0 & -I} [\star]^\top\\
& = \bmat{I & A & B}
\bmat{I & x(1) & \dots & x(T)\\ 0 & -x(0) & 
\dots & -x(T-1) \\ 0 & -u(0) & \dots & -u(T-1)} \cdoT
\bmat{T\epsilon I & 0\\ 0 & -I}
[\star]^\top \\
& = \sum_{i=0}^{T-1} \Big( \epsilon I - (x(i+1)- A x(i) - B u(i)) \cdoT [\star]^\top \Big).
\end{align*}
\endgroup
\end{small}%
Each term of the sum is positive semidefinite since $(A,B)$ is 
assumed to belong to $\bigcap_{i=0}^{T-1} \mathcal{C}_i$, 
hence the whole sum is positive semidefinite. 
This proves that $(A,B) \in \mathcal{C}$.
\end{proof}

Proposition~\ref{prop:set I included in set C} establishes that the set $\mathcal{I}$ is contained in, or at most equal to, 
the set $\mathcal{C}$. We are going to show now through numerical 
experiments that the former set actually has size significantly smaller than the latter.

\section{Comparison between energy and instantaneous bounds: Numerical evidence}
\label{sec:num example}

In this section, we complement the theoretical results in 
Propositions~\ref{prop:feas implication} and \ref{prop:set I included in set C} 
with numerical evidence showing the actual relation between $\mathcal{I}$ 
and $\mathcal{C}$, their dependence on $T$, and the impact of these two aspects on feasibility of the control design problems. 
For this comparison, we preliminarily obtain in Section~\ref{sec:overapprox overline I set} 
an over-approximation of $\mathcal{I}$. We note that this over-approximation, 
although providing insights on this comparison, does not play any role in controller design.

\subsection{Overapproximation $\overline{\mathcal{I}}$ of the set $\mathcal{I}$}
\label{sec:overapprox overline I set}

Whereas the size of $\mathcal{C}$ can be obtained as in 
Section~\ref{sec:sizes of matrix ell}, a measure of 
$\mathcal{I}:=\bigcap_{i=0}^{T-1} \mathcal{C}_i$ is difficult to obtain exactly. 
In fact, $\mathcal{I}$ is an intersection of matrix ellipsoids, and 
for $n=m=1$, even finding the ellipsoid of minimum volume containing $\mathcal{I}$ is NP-complete \cite[p.~44]{boyd1994linear}. 
We thus set up a convex optimization problem to obtain a computable over-approximation $\overline{\mathcal{I}}$ of $\mathcal{I}$. 
Considering $\overline{\mathcal{I}}$ for the comparison with $\mathcal{C}$ is relevant because, albeit being an over-approximation, its size is still (significantly) smaller than the size of $\mathcal{C}$ in all the subsequent numerical examples.
 
\begin{figure}
\centerline{\includegraphics[scale=.75]{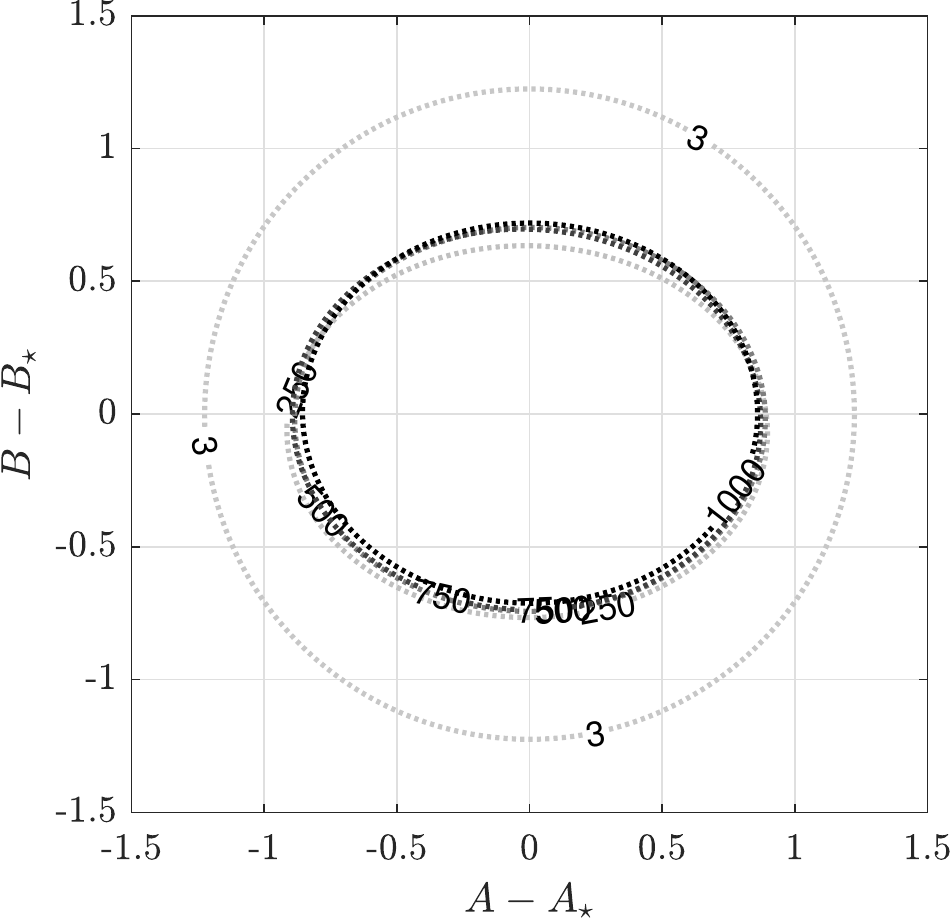}}
\caption{
The ellipsoids inside the dashed lines correspond to $\mathcal{C}$ for $T=3,250,500,750,1000$, as the labels indicate. The larger $T$, the darker the dashed line.}
\label{fig:ellipsoids energy}
\end{figure}
 
We assume that the set $\mathcal{I}$ is bounded, cf. Section \ref{sec:setsC}.
Using the approach in~\cite[\S 3.7.2]{boyd1994linear} for classical ellipsoids, 
we find a matrix ellipsoid $\overline{\mathcal{I}}$ that includes all 
$\mathcal{C}_i$'s through the lossy S-procedure in Lemma~\ref{lemma:lossy S-proc}, 
and we minimize the size of $\overline{\mathcal{I}}$, defined in 
Section~\ref{sec:sizes of matrix ell}. We take $\overline{\mathcal{I}}$ as
\begingroup
\setlength\arraycolsep{2.5pt}
\thinmuskip=.8mu plus 1mu
\medmuskip=1.6mu plus 2mu
\thickmuskip=2.4mu plus 3mu
\begin{align*}
\overline{\mathcal{I}}
& :=\{ (A,B) \colon 
\big[
\begin{array}{@{}c|cc@{}}
I & A & B
\end{array}
\big]
\bmat{
\overline{\sa{C}}~ & \overline{\sa{B}}^\top\\
\overline{\sa{B}}~ & \overline{\sa{A}}}
\left[
\begin{array}{@{}c@{}}
I \\ \hline 
\phantom{\rule{.1pt}{10pt}} A^\top\\
B^\top
\end{array}
\right] \preceq 0
\} \\
& = \{ (A,B) \colon 
\overline{\sa{C}} + \overline{\sa{B}}^\top 
\bmat{
A^\top \\
B^\top
}
+ 
\bmat{A & B} \overline{\sa{B}}
+
\bmat{A & B} \overline{\sa{A}} \bmat{
A^\top \\
B^\top
} \preceq 0
\}
\end{align*}
\endgroup
where we set $\overline{\sa{C}}=\overline{\sa{B}}^\top \overline{\sa{A}}^{-1} \overline{\sa{B}} - I$ (otherwise the representation of $\overline{\mathcal{I}}$ is homogeneous) and require $\overline{\sa{A}} \succ 0$ so that \eqref{ell matr assump} holds. 
We impose $\mathcal{I}=\bigcap_{i=0}^{T-1} \mathcal{C}_i 
\subseteq \overline{\mathcal{I}}$ via the lossy matrix S-procedure in Lemma~\ref{lemma:lossy S-proc}, and obtain
\begingroup
\setlength\arraycolsep{2.5pt}
\begin{align*}
\bmat{
\overline{\sa{B}}^\top \overline{\sa{A}}^{-1} \overline{\sa{B}}\! -\! I & \,\overline{\sa{B}}^\top\\
\overline{\sa{B}} & \,\overline{\sa{A}}}-
\sum_{i =0}^{T-1}
\tau_i
\bmat{\sa{C}_i	& \sa{B}_i^\top\\
\sa{B}_i		& \sa{A}_i} \preceq 0, \tau_i \ge 0 \text{ for } i \in \mathbb{I}
\end{align*}
\endgroup
with $\sa{C}_i$, $\sa{B}_i$, $\sa{A}_i$ defined in~\eqref{ABC_C_Ci}. 
Since the first inequality is nonlinear in the decision variables 
$\overline{\sa{B}}$ and $\overline{\sa{A}}$, we rewrite it by Schur complement as
\begin{equation}
\label{bigContainCond}
\bmat{
\vspace*{1pt}
- I - \sum_{i =0}^{T-1} \tau_i\sa{C}_i   &
\overline{\sa{B}}^\top - \sum_{i =0}^{T-1} \tau_i\sa{B}_i^\top &
\overline{\sa{B}}^\top\\
\vspace*{1pt}
\overline{\sa{B}} - \sum_{i =0}^{T-1} \tau_i\sa{B}_i &
\overline{\sa{A}} - \sum_{i =0}^{T-1} \tau_i\sa{A}_i &
0\\
\overline{\sa{B}} &
0 &
- \overline{\sa{A}}
} \preceq 0.
\end{equation}
By~\eqref{size n>=1}, the size of $\overline{\mathcal{I}}$ is 
given by $(\det \overline{\sa{A}} )^{-\frac{n}{2}}$ thanks to the adopted normalization. 
These constraints and objective function result in the optimization problem
\begin{equation}
\label{overapprox set I}
\begin{aligned}
& \text{minimize} & & -\log\det \overline{\sa{A}} \\
& \text{subject to} & & \eqref{bigContainCond},\, \overline{\sa{A}} \succ 0, \, 
\tau_i\ge 0 \text{ for } i \in \mathbb{I}.
\end{aligned}%
\end{equation}
When $n=m=1$, this optimization problem boils down to 
that in~\cite[Eq.~(3.15)]{boyd1994linear}, but \eqref{overapprox set I} 
also captures the case of generic dimensions $n$ and $m$, 
as we need in the following. 

\subsection{A visualizable example}
\label{sec:energy vs ampl:visual example}

In this section we examine more thoroughly the system 
of Example~\ref{ex:noMonoEx} and show how the sizes 
of $\mathcal{C}$ and $\mathcal{I}$ depend on $T$ and 
how they compare with each other.

\begin{figure}
\centerline{\includegraphics[scale=.75]{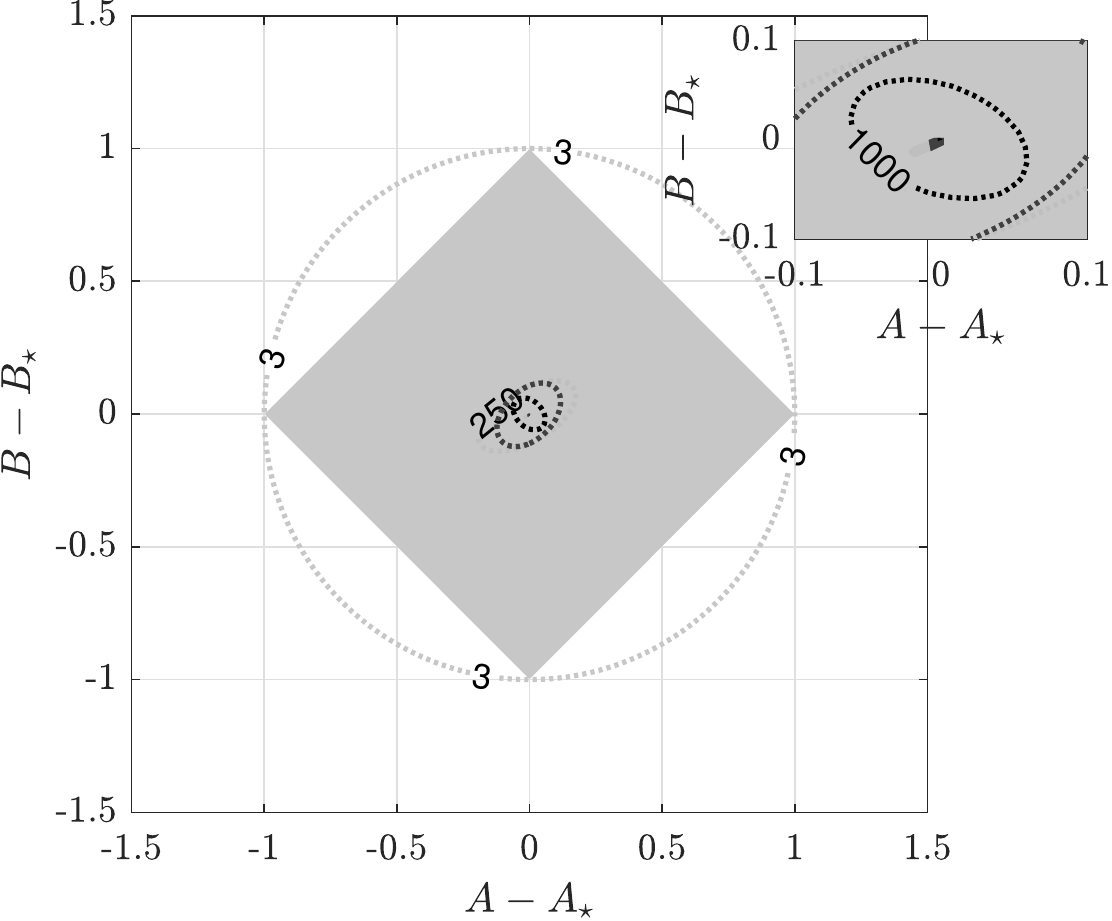}}
\caption{
The shaded areas correspond to $\mathcal{I}$ and are barely visible for $T\ge 250$, as can be appreciated in the inset.
The ellipsoids inside the dashed lines correspond to $\overline{\mathcal{I}}$ (we omit some labels for readability). Increasingly darker color for the areas and lines correspond to $T=3,250,500,750,1000$.}
\label{fig:ellipsoids ampl}
\end{figure}

We prolong the data sequences of Example~\ref{ex:noMonoEx} by using as input and disturbance the realizations of random variables uniformly distributed in $[-2,2]$ and $[-{\epsilon},{\epsilon}]=[-1,1]$, respectively.
The resulting set $\mathcal{C}$, depicted for $T=3$ in Fig.~\ref{fig:noMonoEx}, is now depicted for some $T$ up to $1000$ in Fig.~\ref{fig:ellipsoids energy}. 
Fig.~\ref{fig:ellipsoids energy} shows that the energy-based ellipsoids $\mathcal{C}$ shrink very slowly as $T$ increases, and stay approximately constant from $T=250$ on.
The resulting set $\mathcal{I}$ is depicted for the same values of $T$ in Fig.~\ref{fig:ellipsoids ampl} together with its over-approximation $\overline{\mathcal{I}}$, determined as in Section~\ref{sec:overapprox overline I set}. 
Fig.~\ref{fig:ellipsoids ampl} shows that the ellipsoids $\overline{\mathcal{I}}$ 
shrink very quickly with $T$ and, albeit only over-approximations of $\mathcal{I}$, they are significantly smaller than the ellipsoids $\mathcal{C}$. 
The actual sets $\mathcal{I}$, which are depicted by shaded areas, shrink even faster to the point that they are barely visible already for $T=250$.

\begin{figure}
\centerline{\includegraphics[scale=.65]{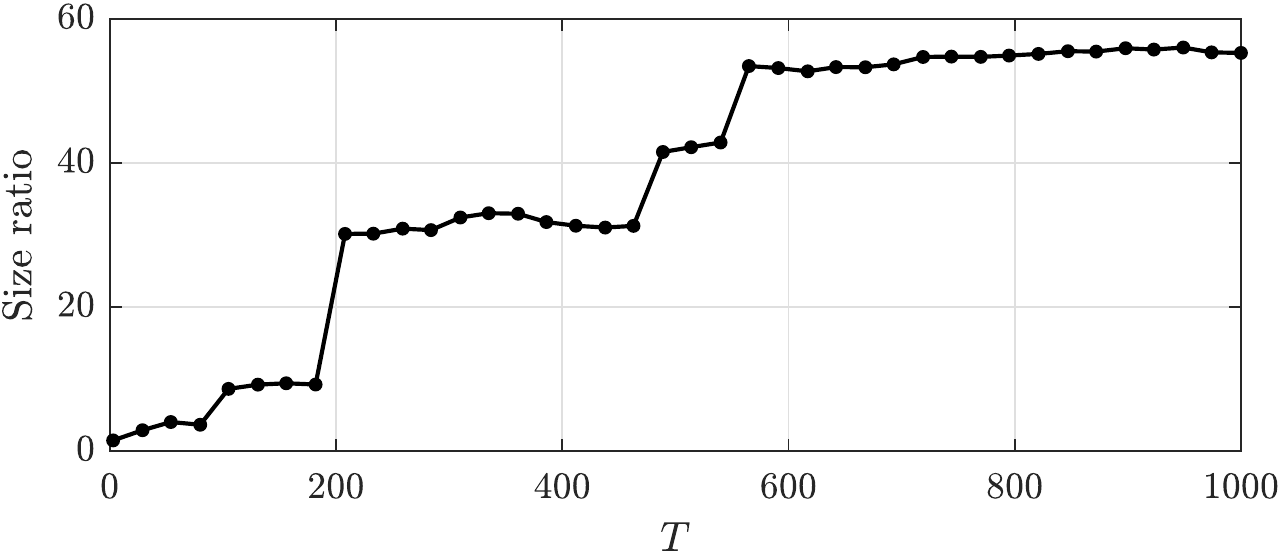}}
\caption{Ratio between the sizes of $\mathcal{C}$ and $\overline{\mathcal{I}}$, Section~\ref{sec:energy vs ampl:visual example}.}
\label{fig:twoSizes}
\end{figure}

For a clearer visualization, Fig.~\ref{fig:twoSizes} depicts 
the ratio between the sizes of $\mathcal{C}$ and $\overline{\mathcal{I}}$. $\overline{\mathcal{I}}$
is smaller than $\mathcal{C}$ by more than 30 times already with 
approximately $200$ data points.

\subsection{Example with third order dynamics}
\label{sec:num example:unstable}

In this section, we delve into the considerations of Section~\ref{sec:energy vs ampl:visual example} through 
a more complex example with matrices
\begingroup%
\setlength\arraycolsep{2.5pt}%
\begin{equation*}
A_\star=\bmat{    
	0.1274 &   0.1431  &  0.1974\\
    0.3619 &   0.6292  &  0.4153\\
    0.6972 &   0.1574  &  0.4111}\!,\,
B_\star=\bmat{
    0.6901 & 0.9047\\
    0.4809 & 0.6030\\
    0.8913 & 0.1478}.
\end{equation*}
\endgroup
A data set is obtained by applying the realization of a 
Gaussian random variable (zero mean and unit standard deviation) as input, and as disturbance the realization of a random variable distributed uniformly in $\{ d \colon |d|^2 \le {\epsilon} \}$ with 
different values for ${\epsilon}$.

\begin{figure}
\centerline{\includegraphics[scale=.65]{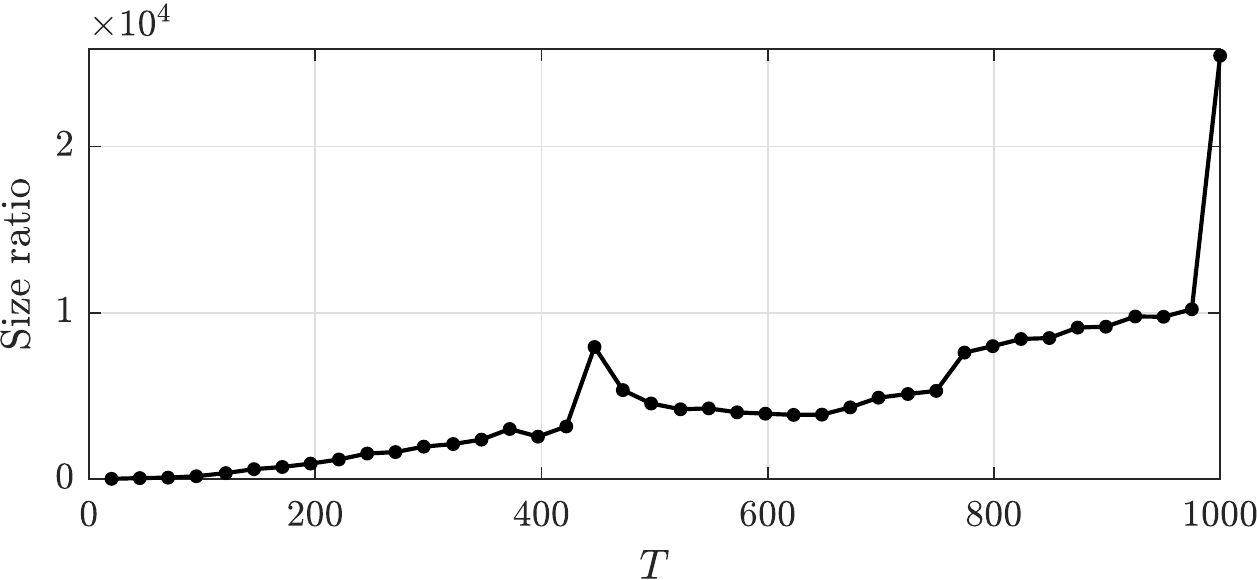}}
\caption{Ratio between the sizes of $\mathcal{C}$ 
and $\overline{\mathcal{I}}$, Section~\ref{sec:num example:unstable}.}
\label{fig:twoSizes3rd}
\end{figure}

\begin{figure}
\centerline{\includegraphics[scale=.65]{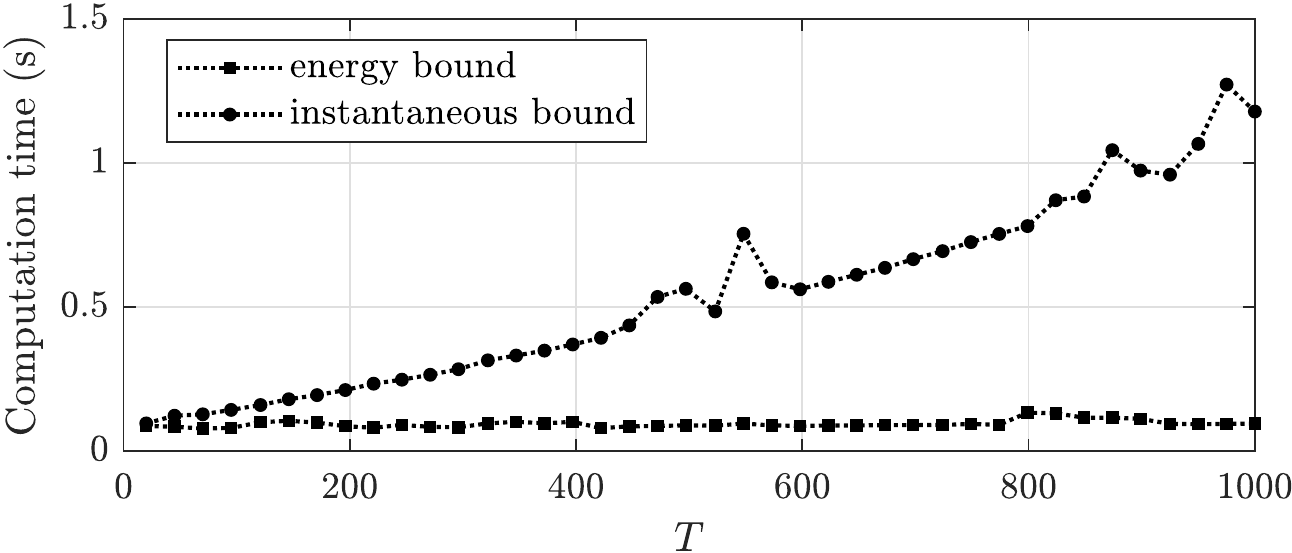}}
\caption{Computation times for controller design in \eqref{energy-based approach <=>} and \eqref{ampl-based approach <=}.}
\label{fig:comptime2appro3rd}
\end{figure}

As in the previous example, we compare the sizes of $\mathcal{C}$ and of $\overline{\mathcal{I}}$ assuming $\epsilon=0.1$. 
Their ratio is plotted in Fig.~\ref{fig:twoSizes3rd} for increasingly larger portions of data from $T=20$ up to $T=1000$.
The size of $\overline{\mathcal{I}}$ is smaller than $\mathcal{C}$ 
by approximately two orders of magnitude already with $100$ data points.
We commented after Proposition~\ref{prop:ampl-based approach <=} 
that the energy-based approach \eqref{energy-based approach <=>} 
requires $T-1$ scalar decision variables less than the formulation 
in~\eqref{ampl-based approach <=}. 
We show numerically that the impact of these extra variables is  modest even with a large number of data. 
The computations times\footnote{
These wall-clock times are obtained through the MATLAB\textsuperscript{\textregistered} 
R2019b function \texttt{timeit}, which automatically executes the 
program multiple times and computes a median, on a machine with processor 
Intel\textsuperscript{\textregistered} Core{\texttrademark} i7 with 4 cores and 1.80~GHz.} 
to solve \eqref{energy-based approach <=>} or \eqref{ampl-based approach <=} are depicted in Fig.~\ref{fig:comptime2appro3rd}.
The controller design in~\eqref{energy-based approach <=>} has the appealing feature that the associated computation time barely changes with the number of data points whereas \eqref{ampl-based approach <=} exhibits a dependence on $T$ that results in a computation time that grows linearly with $T$.

\begin{figure}
\centerline{\includegraphics[scale=.65]{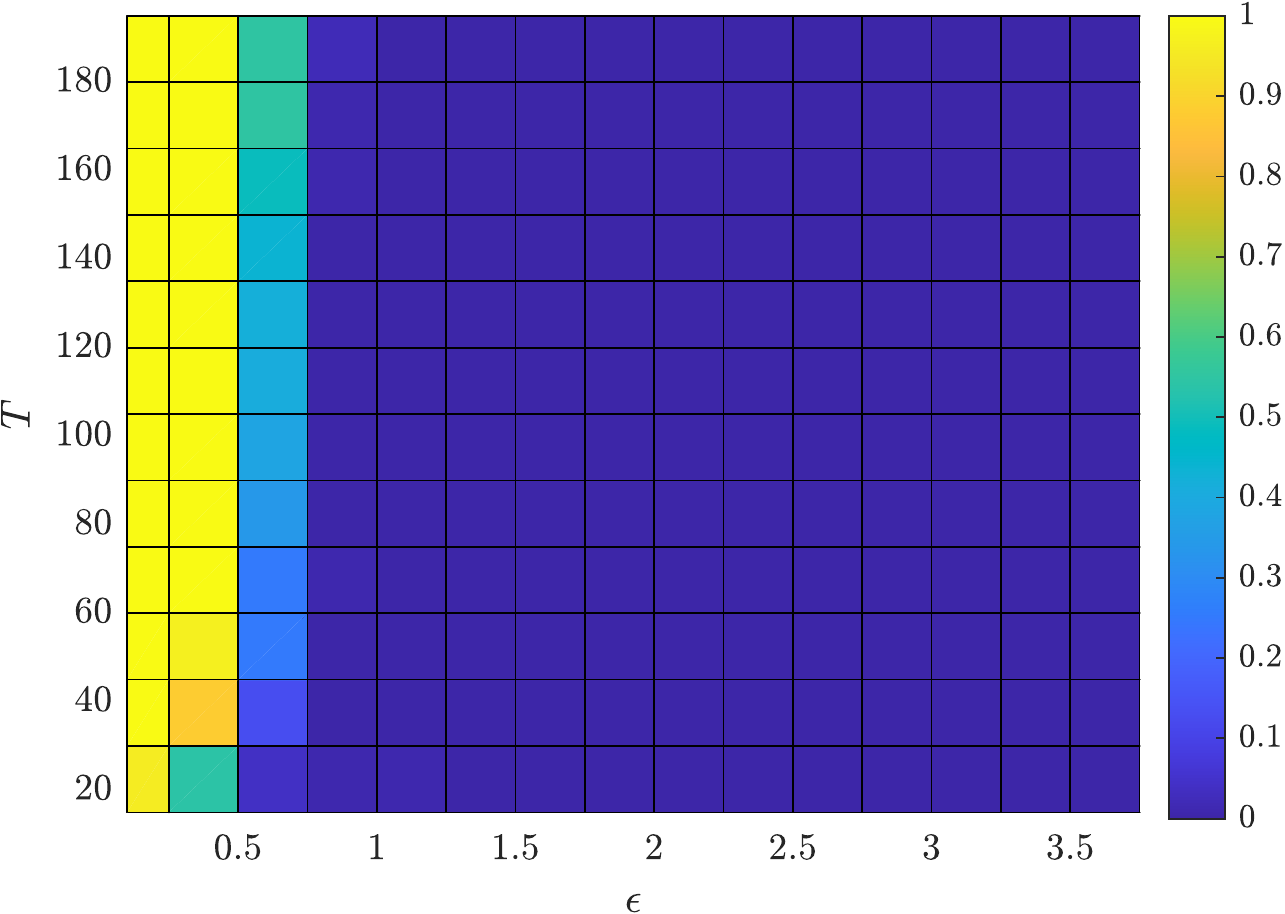}}
\caption{Energy-bound approach \eqref{energy-based approach <=>}: the color represents for each disturbance bound $\epsilon$ 
and data length $T$ the ratio of feasible controller design problems in a batch of 100.}
\label{fig:feasperc2appro3rd_en}
\end{figure}

\begin{figure}
\centerline{\includegraphics[scale=.65]{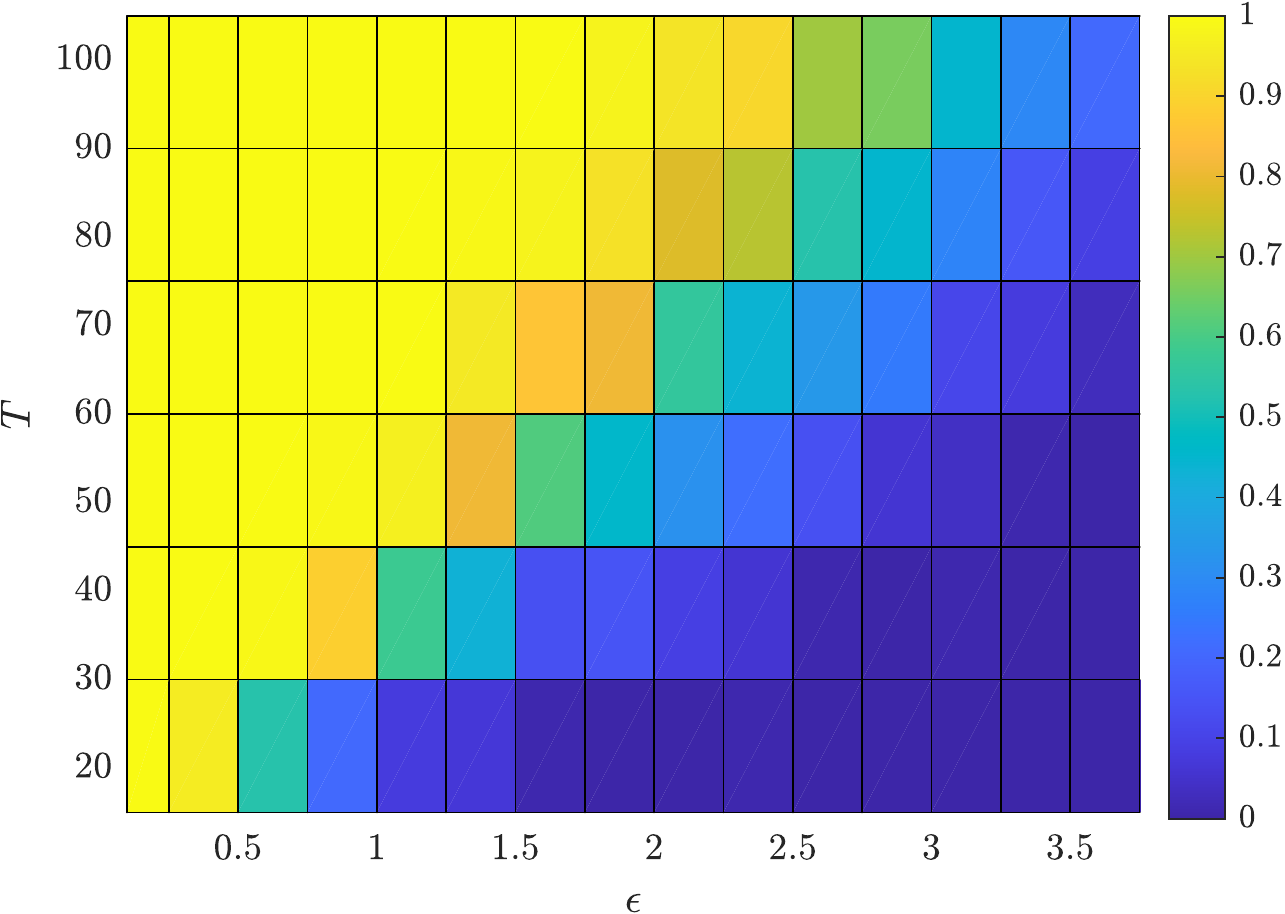}}
\caption{Instantaneous-bound approach \eqref{ampl-based approach <=}, see caption of Fig.~\ref{fig:feasperc2appro3rd_en}.}
\label{fig:feasperc2appro3rd_am}
\end{figure}

Finally, we examine the joint effect of the disturbance bound 
$\epsilon$ and of the length $T$ on both the approaches.
For each value of $\epsilon$ and $T$, we solve a batch of 100 feasibility problems \eqref{energy-based approach <=>} or \eqref{ampl-based approach <=}, count the number $n_\textup{feas}$ of feasible ones in that batch and display the ratio $n_\textup{feas}/100\in[0,1]$ in Figs.~\ref{fig:feasperc2appro3rd_en}--\ref{fig:feasperc2appro3rd_am} through a color code. Yellow areas are the good ones in terms of feasibility. 
Fig.~\ref{fig:feasperc2appro3rd_en} depicts this ratio for the energy-bound approach and shows that above a threshold $\epsilon$ of approximately $0.7$, the problem \eqref{energy-based approach <=>} has no solutions regardless of the number of data. 
This behaviour seems consistent with what we observed in Fig.~\ref{fig:ellipsoids energy} about the fact that the set of consistent matrices is not reduced by larger amounts of data.
Fig.~\ref{fig:feasperc2appro3rd_am} depicts the ratio $n_\textup{feas}/100$ for the instantaneous-bound approach, and shows its appealing feature of being able to withstand increasingly larger disturbance magnitudes as long as an increasingly number of data points is collected to reduce uncertainty.
Comparing Figs.~\ref{fig:twoSizes3rd}--\ref{fig:comptime2appro3rd} 
with Figs.~\ref{fig:feasperc2appro3rd_en}--\ref{fig:feasperc2appro3rd_am}, we can conclude that the price paid in terms of computation time is negligible compared with the gain in terms of robustness.

\section{Conclusions}
\label{sec:concl}

In this paper, we compared two different disturbance models 
for data-driven control design: a model $\mathcal{D}_\tu{e}$ considering energy bounds on the entire disturbance sequence
and a model $\mathcal{D}_\tu{i}$ considering instantaneous bounds on the disturbance. 
The model $\mathcal{D}_\tu{e}$ leads to elegant necessary and 
sufficient condition for stability, and results in a design program
whose number of variables depends only on the dimensions of the 
system to be controlled and not on the number of data points. 
On the other hand, the model $\mathcal{D}_\tu{i}$
reflects a much more natural way of describing disturbances.

We analyzed pros and cons of working directly with $\mathcal{D}_\tu{i}$, instead of converting it into $\mathcal{D}_\tu{e}$.
The analysis shows that, computational considerations aside, it is in fact {always} preferable to work with $\mathcal{D}_\tu{i}$.
First, $\mathcal{D}_\tu{i}$ results in a design program that is 
always feasible whenever the one associated with $\mathcal{D}_\tu{e}$ is feasible. 
Second, numerical evidence shows that the feasibility gap can be extremely large. 
This latter aspect has been analyzed by introducing a notion of \emph{size} for the uncertainty set induced by the disturbance.
Simulations show that while the uncertainty set associated with $\mathcal{D}_\tu{e}$ does not necessarily shrink, the
one associated with $\mathcal{D}_\tu{i}$ often shrinks very quickly, and is many order of magnitudes smaller. 
As for computational considerations, working with $\mathcal{D}_\tu{i}$ is less advantageous because it results in a design program whose number of variables depends on the number of data points. 
With the exception of extremely large data sets, however, the price paid in terms of computation time appears negligible compared with the advantages offered in terms of robustness.

\bibliographystyle{plain}
\bibliography{references}

\end{document}